\documentstyle[12pt]{article}
\font \fivesans               = cmss10 at 5pt
\font \sevensans              = cmss10 at 7pt
\font \tensans                = cmss10
\newfam\sansfam
\textfont\sansfam=\tensans\scriptfont\sansfam=\sevensans
\scriptscriptfont\sansfam=\fivesans
\def\sans{\fam\sansfam\tensans}
\def\Operator#1{\mathchoice
   {\mbox{\boldmath $#1$}}{\mbox{\boldmath $#1$}}
   {\mbox{\footnotesize \boldmath $#1$}}
   {\mbox{\footnotesize \boldmath $#1$}}}
\def\bbbz{{\mathchoice {\hbox{$\sans\textstyle Z\kern-0.4em Z$}}
{\hbox{$\sans\textstyle Z\kern-0.4em Z$}}
{\hbox{$\sans\scriptstyle Z\kern-0.3em Z$}}
{\hbox{$\sans\scriptscriptstyle Z\kern-0.2em Z$}}}}
\font\boldit=cmbxti10 scaled \magstep1
\def\Notion#1{{\boldit #1\/}}
\def\rstop#1{\right.}
\def\bbbr{{\rm I\!R}}

\def\supp{\mbox{\rm supp\,}}
\def\jv{{\vec \jmath}}
\def\Nv{{\vec \nabla}}
\def\OA{{\Operator A}}
\def\Op#1{{\Operator #1}}
\def\xv{{\vec x}}
\def\modulus#1{\left| #1 \right|}
\def\norm#1{\left\| #1 \right\|}

\begin{document}

\title{\hfill {\normalsize ASI-TPA/14/97}\linebreak\vskip 2mm
Gisin Nonlocality of the Doebner-Goldin 2-Particle Equation}
\author{W.~L\"{u}cke\thanks{E-mail: aswl@pt.tu-clausthal.de}
\\Arnold Sommerfeld Institute for
Mathematical Physics\\Technical University of Clausthal \\ D-38678
Clausthal, Germany}
\date{October 10, 1997}
\maketitle
\begin{abstract}
Gisin's argument against deterministic nonlinear Schr\"odinger equations
is shown to be valid for every (formally) nonlinearizable case of the
general Doebner-Goldin 2-particle equation in the following form:

The time-dependence of the position probability distribution of a
particle `behind the moon' may be instantaneously changed by an
arbitrarily small instantaneous change of the potential `inside the
laboratory'.

\end{abstract}

\textheight 22.5cm
\section{Introduction}

Several years ago N.~Gisin pointed out that for  every
(deterministic, scalar) nonlinear 2-particle Schr\"odinger theory
there is an initial wave function $\Psi_0(\xv_1,\xv_2)\in
L^2(\bbbr^3)\otimes L^2(\bbbr^3)$ and a self-adjoint operator
$\OA\otimes \Op 1$ such that the `expection value' of  $\OA\otimes \Op
1$ may be almost instantaneously influenced by performing measurements on
particle 2 \cite{Gisin2,Gisin3}. As shown in \cite{LueckeNL} the
existence of such a \Notion{Gisin effect} does not depend on
Gisin's questionable assumptions concerning the measuring process. More
precisely, the following holds:
\vskip 5mm

There is an initial wave function $\Psi_0(\xv_1,\xv_2)\in
L^2(\bbbr^3)\otimes L^2(\bbbr^3)$ and a self-adjoint operator
$\OA\otimes \Op 1$ such that
$$
\left\langle \Psi^{V_t}_t\mid\OA\otimes \Op 1\,\Psi^{V_t}_t  \right\rangle
$$
depends nontrivially on $V_t\,$, where $ \Psi^{V_t}_t$ denotes the solution of
the corresponding initial value problem for the nonlinear 2-particle
Schr\"odinger equation\footnote{We use units in which $\hbar=1$
and $m=1\,$.}
\begin{equation} \label{GNLS}
i\partial_t\Psi_t(\xv_1,\xv_2) = \left(-\frac{1}{2} \Delta +
V_t(\xv_2)\right) \Psi_t(\xv_1,\xv_2)
+F_t[\Psi_t](\xv_1,\xv_2)\;,\quad\Delta=\Delta_(\xv_1,\xv_2)\,,
\end{equation}
even if the latter is formally local in the sense that  the nonlinearity
$F_t$
is a {\bf local} (non-linear) functional:
$$
F_t[\Psi](\xv_1,\xv_2) = F_t[\Phi](\xv_1,\xv_2)
\quad\forall\,(\xv_1,\xv_2) \notin\supp(\Psi-\Phi)\,,
$$
\vskip 5mm

\noindent
However, in a nonlinear quantum theory one cannot consider all linear
self-adjoint operators as physical observables. Therefore such
Gisin effects may be completely irrelevant as in the linearizable case
of the general Doebner-Goldin equation \cite{LueckeNL}. In order to
stress this we call a Gisin effect \Notion{relevant} if the
corresponding operator $\OA\otimes\Op 1$ is a physical observable. If
$\OA$ is a multiplier in $L^2(\bbbr^3)$ this is certainly the case due
to the fundamental assumption of nonlinear quantum mechanics:
$$
\modulus{\Psi_t(\xv_1,\xv_2)}^2 = \left\{\begin{array}[c]{l}
\mbox{probability density for localization of particle 1 around } \xv_1\\
\mbox{and particle 2 around }\xv_2 \mbox{ at time }t\,.
\end{array}\rstop\}
$$
\vskip 5mm

The purpose of the present paper is to show that there are relevant
Gisin effects for every case, except $c_2= -2c_5\,$, of the 2-particle
\Notion{general Doebner-Goldin equation} \cite{DoGo,DoGoGen}. Using the
short-hand notation
$$
\rho_t \stackrel{\rm def}{=} \modulus{\Psi}^2\;, \quad \jv_t
\stackrel{\rm def}{=} \Im \left(\overline{\Psi_t} \Nv \Psi_t\right)\;,
\quad \Nv = \Nv_{(\xv_1,\xv_2)}\,,
$$
this nonlinear Schr\"odinger equation,\footnote{For simplicity
we consider only the special case $V_t(\xv_1,\xv_2) = V_t(\xv_2)\,$.} 
up to some {\em nonlinear gauge transformation}
$$
\Psi \longmapsto e^{i\lambda
\ln\modulus{\Psi}} \Psi\;,\quad\lambda\in\bbbr\,,
$$
is given by (\ref{GNLS}) and
\begin{equation} \label{DG}
F[\Psi] = \left(c_1 \frac{\Nv\cdot
\jv}{\rho} + c_2 \frac{\Delta\rho}{\rho} + c_3 \frac{\jv\,\,^2}{\rho^2}+
c_4 \frac{\jv\cdot \Nv\rho}{\rho^2} + c_5
\frac{(\Nv\rho)^2}{\rho^2}\right) \Psi\,,
\end{equation}
with real parameters $c_1,\ldots,c_5$ \cite{DoGoPR,DoGoNa}.
\medskip

\section{Previous Results}

We assume that there are sufficiently many, well-behaved,
\hbox{$V_t$-dependent} solutions of (\ref{GNLS}),(\ref{DG}) -- at least
locally in time -- which are physically acceptable in the following
sense:\footnote{This assumption is well known to be fulfilled for the
linear Schr\"odinger equation and should follow along the lines
discussed in \cite{Teismann} for the Doebner-Goldin equation.}
\begin{quote}
Switching on  $V_t$ instantaneously does not cause an instantaneous
change of the wave function.
\end{quote}
Then we have a relevant Gisin effect whenever there is a sufficiently
well-behaved initial wave function $\Psi_0$ and some integer
$k$ for which the function
\begin{equation} \label{instGE}
\left(\left(\frac{\partial}{\partial t}\right)^k \int
\rho_t(\xv_1,\xv_2) \,{\rm d}\xv_2\right)_{|_{t=0}}
\end{equation}
of $\xv_1$ depends nontrivially on $V_t=V\,$. That such instantaneous
Gisin effects exist unless
\begin{equation} \label{Werner}
c_3=c_1+c_4=0
\end{equation}
was first shown by R.\ Werner \cite{Werner}.
Since Werner's Ansatz, using entangled Gau\ss{} solutions for oscillator
potentials, was too special the $V$-dependent part of (\ref{instGE}) was
calculated in \cite{LuNa} for general initial conditions and $k\leq
3\,$. This only confirmed Werner´s result. We will show, however, that
(\ref{Werner}) does not exclude nontrivial $V$-dependence of
(\ref{instGE}) for $k=4\,$.
\medskip

\section{Additional Gisin Effects}

Since (\ref{instGE}) becomes very complicated for $k>3$ we calculate
$
\left(\left(\frac{\partial}{\partial t}\right)^k \int x_1^1
\rho_t(\xv) \,{\rm d}\xv\right)_{|_{t=0}}\,,
$
instead, assuming $\Psi_t$ to be sufficiently well behaved.
\vskip 5mm

Using the continuity equation
$$
\partial_t \rho_t + \Nv\cdot\jv_t = 0
$$
we get by partial integration w.r.t.\ $x_1^1$ the first Ehrenfest
relation
$$
\partial_t\int x_1^1\,\rho_t(\xv)\,{\rm d}\xv = \Im \int
\overline{\Psi_t(\xv)}\,\partial_1 \Psi_t(\xv) \,{\rm d}\xv\,,
$$
where
$$
\partial_t\stackrel{\rm def}{=}\frac{\partial}{\partial t}\;,\quad
\partial_1 \stackrel{\rm def}{=}\frac{\partial}{\partial x_1^1}\;,
\quad \xv \stackrel{\rm def}{=}(\xv_1,\xv_2)\,.
$$
Further differentiation w.r.t.\ $t$ yields
$$
\left(\partial_t\right)^2 \int x^1\,\rho_t(\xv)\,{\rm d}\xv = -\int
\rho_t(\xv)\, \partial_1 R[\Psi_t](\xv)\,{\rm d}\xv
$$
(note that since $\partial_1 V=0$),
\begin{equation} \label{ess3}
\left(\partial_t\right)^3 \int
x^1\,\rho_t(\xv)\,{\rm d}\xv = \int \Nv\cdot\jv(\xv,t)\,\partial_1
R[\Psi_t](\xv)\,{\rm d}\xv - \int
\rho_t(\xv)\, \partial_1 \partial R[\Psi_t](\xv)\,{\rm d}\xv\,,
\end{equation}
and finally
\begin{equation} \label{ess4}
\begin{array}[c]{rcl}
\displaystyle \left(\partial_t\right)^4 \int x^1\,\rho_t(\xv)\,{\rm
d}\xv &=& \displaystyle \int \left( \partial_t\Nv\cdot \vec\jmath_t(\xv)
\right)\,\partial_1 R[\Psi_t](\xv)\,{\rm d}\xv\\
&&\displaystyle\phantom{y}+ 2\int \left( \Nv\cdot \vec\jmath_t(\xv)
\right)\,\partial_1\partial_t R[\Psi_t](\xv)\,{\rm d}\xv\\
&&\displaystyle \phantom{y} - \int \rho_t(\xv)\, \partial_1\partial_t^2
R[\Psi_t](\xv)\,{\rm d}\xv\,,
\end{array}
\end{equation}
where
$$
F[\Psi_t](\xv) = R[\Psi_t](\xv)\,\Psi_t(\xv)\,.
$$
Since the $V$-dependent part of $\left(\partial_t \Nv\cdot 
\vec\jmath_t(\xv)\right)_{|_{t=0}}$ is\footnote{We always denote by $
{\rm ess\,}(X)$ the partial sum of  $V$-dependent items of $X\,$.}
\begin{equation} \label{ess-dot-diff-j}
\begin{array}[c]{rcl}
\displaystyle {\rm ess\,}\left(\partial_t \Nv\cdot
\vec\jmath_t(\xv)\right)_{|_{t=0}} &=&\displaystyle
\Nv\cdot\Re\left(\overline{\Psi_0(\xv)}[V(\xv_2),\Nv]_-
\Psi_0(\xv)\right)\\
&=& \displaystyle -\Nv\cdot\left( \rho_0(\xv)\Nv V(\xv_2)\right)\,,
\end{array}
\end{equation}
(\ref{ess3}) and (\ref{ess4}) depend linearly on $R$ and therefore the
contributions of the different terms in (\ref{DG}) may be checked
separately.
\vskip 5mm

\noindent
Let us first consider the special case $c_4=-c_1\ne 0\,,\;c_\nu=0\,$
else, i.e.
\begin{equation} \label{case1}
R[\Psi](\xv) = c_1\left( \frac{\Nv\cdot \jv}{\rho} - \frac{\jv\cdot
\Nv\rho}{\rho^2}\right) = c_1\,\Delta \arg\left(\Psi(\xv)\right)\,.
\end{equation}
Here
$$
\partial_t R[\Psi_t](\xv) = -c_1\Delta \Re\left( \frac{i\partial_t
\Psi_t(\xv)}{\Psi_t(\xv)}\right)
$$
and therefore
$$
{\rm ess\,}\left(\partial_tR[\Psi_t](\xv)\right)_{|_{t=0}} =
-c_1\,\Delta V(\xv_2)\,.
$$
Due to
$$
\begin{array}[c]{rcl}
\partial_1{\rm ess}\left(\partial_t^2 R[\Psi_t(\xv)]\right)_{|_{t=0}}
&=& \displaystyle \frac{c_1}{2}\,\partial_1\Delta \, {\rm ess}\left(
\Re \left( \partial_t \frac{\Delta \Psi_t(\xv)}{\Psi_t(\xv)} \right)
\right)_{|_{t=0}}\\
&=& \displaystyle \frac{c_1}{2}\,\partial_1\Delta \,
\Im\left(\frac{[\Delta,V(\xv_2)]_-\Psi_0(\xv)}{\Psi_0(\xv)} \right)\\
&=& -c_1\,\partial_1\Delta \left(\left(\Nv V(\xv_2)\right)\cdot\Nv
\arg\left(\Psi_0(\xv)\right)\right)
\end{array}
$$
the $V$-dependent part of (\ref{ess4}) at $t=0$ is
\begin{equation} \label{result1}
\begin{array}[c]{l}
\displaystyle {\rm ess}\left(\left(\partial_t\right)^4
\int x^1\,\rho_t(\xv)\,{\rm d}\xv\right)_{|_{t=0}}\\
\displaystyle = -c_1\int \left(\Nv\cdot\left(\rho_0(\xv)\Nv
V(\xv_2)\right)\right) \Delta\partial_1 \arg(\Psi_0(\xv))\,{\rm d}\xv\\
\displaystyle \phantom{=} -c_1\int \rho_0(\xv)\,\partial_1
\Delta\left( (\Nv V(\xv_2))\cdot \Nv \arg(\Psi_0(\xv))\right)\,{\rm
d}\xv\\
\displaystyle = -c_1\int \rho_0(\xv)\, [\Delta,\Nv V(\xv_2)]_- \cdot \Nv
\partial_1 \arg(\Psi_0(\xv))\,{\rm d}\xv\,.
\end{array}
\end{equation}
Obviously, (\ref{result1}) does not always vanish. Hence there are
relevant Gisin effects for (\ref{case1}).
\vskip 5mm

\noindent
Next, let us consider the case $c_2\ne 0\,,\; c_\nu=0\,$ else,  i.e.
\begin{equation} \label{case2}
R[\Psi](\xv) = c_2\,\frac{\Delta \rho(\xv)}{\rho(\xv)}\,.
\end{equation}
Then, since
$$
{\rm ess}\left(\partial_t^2 \frac{\Delta
\rho_t(\xv)}{\rho_t(\xv)}\right)_{|_{t=0}} =\left(\frac{\Delta
\rho_0(\xv)}{\rho_0(\xv)^2} -\frac{1}{\rho_0(\xv)}\Delta\right) {\rm
ess\,}\left(\partial_t \Nv\cdot \vec\jmath_t(\xv)\right)_{|_{t=0}}\,,
$$
the $V$-dependent part of (\ref{ess4}) at $t=0$ is
$$
\begin{array}[c]{l}
\displaystyle {\rm ess}\left(\left(\partial_t\right)^4
\int x^1\,\rho_t(\xv)\,{\rm d}\xv\right)_{|_{t=0}}\\
= \displaystyle  -c_2\int \left(\Nv\cdot\left(\rho_0(\xv)\Nv
V(\xv_2)\right)\right) \partial_1
\frac{\Delta \rho_0(\xv)}{\rho_0(\xv)}\,{\rm d}\xv\\
\displaystyle\phantom{=}+c_2 \int \rho_0(\xv)\partial_1\left(\left(
\frac{\Delta \rho_0(\xv)}{\rho_0(\xv)^2} -\frac{1}{\rho_0(\xv)}
\Delta\right)\left( \Nv\cdot\left(\rho_0(\xv)\Nv
V(\xv_2)\right)\right)\right){\rm d}\xv\\
= \displaystyle c_2\int \left(\left[\Delta,
\frac{1}{\rho_0(\xv)}\right]_-
\partial_1\rho_0(\xv)\right)
\Nv\cdot\left(\rho_0(\xv)\Nv
V(\xv_2)\right){\rm
d}\xv\,.
\end{array}
$$
This is nonzero, for example, when
$$
V(\xv_2) = (x_2^1)^3\;,\quad\rho_0(\xv)= \exp\left(-\norm{\xv}^2 -
x_1^1\,x_2^1\right)\,.
$$
Hence there are relevant Gisin effects for (\ref{case2}), too.
\vskip 5mm

Finally, let us consider the case\footnote{Note that
(\ref{GNLS}),(\ref{DG}) is (formally) linearizable in this case
\cite{NattT93}.}
\begin{equation} \label{final}
c_1=c_3=c_4=0=c_2+2c_5\,,
\end{equation}
i.e.
\begin{equation} \label{ec}
R[\Psi](\xv) = c_2\left(\frac{\Delta \rho(\xv)}{\rho(\xv)} -\frac 12
\left(\frac{\Nv \rho(\xv)}{\rho(\xv)}\right)^2\right)\,.
\end{equation}
In this case (\ref{GNLS}),(\ref{DG}) fulfills also the second Ehrenfest
relation \cite{NattD}. Then already $\left(\partial_t\right)^2 \int
x^1\,\rho_t(\xv)\,{\rm d}\xv$ vanishes for all $t\,$, hence (\ref{ec})
does {\bf not} contribute to (\ref{ess4}).
\medskip

\section{Conclusion} \label{S-Concl}

We already know from \cite{Werner,LuNa} --- or easily rederive from
(\ref{ess3}) --- that for (\ref{DG}) there are relevant Gisin effects
whenever Werner's condition
$$
c_3=c_1+c4=0
$$
is violated. If this condition is fulfilled, $R[\Psi] = F[\Psi]/\Psi$
may be written in the form
$$
\begin{array}[c]{rcl}
R &=& c_1\, \left( R_1 -R_4\right) + c_2\,R_2 + c_5\,R_5\\
&=& c_1\, \left( R_1 -R_4\right) + (c_2+2c_5)\,R_2 + c_5\,\left(-2R_2 +
R_5\right)\,.
\end{array}
$$
where
$$
R_1[\Psi]\stackrel{\rm def}{=} \frac{\Nv\cdot 
\jv}{\rho}\;, \quad R_2[\Psi]\stackrel{\rm def}{=}
\frac{\Delta\rho}{\rho}\;, \quad R_4[\Psi]\stackrel{\rm def}{=}
\frac{\jv\cdot \Nv\rho}{\rho^2}\;, \quad R_5[\Psi]\stackrel{\rm def}{=}
\frac{(\Nv\rho)^2}{\rho^2}\,.
$$
As shown in the previous section, the contributions to ${\rm
ess}\left(\left(\partial_t\right)^4 \int x^1\,\rho_t(\xv)\,{\rm
d}\xv\right)_{|_{t=0}}$ by $R_1-R_4$ and $R_2$ are functionally
independent while $-2R_2 +R_5$ does not contribute. We conclude that
there are relevant instantaneous Gisin effects for the general
Doebner-Goldin equation (\ref{GNLS}),(\ref{DG}) whenever condition
(\ref{final}) is violated.  Due to translation invariance this
represents a serious locality problem:
\begin{quote}
The time-dependence of the position probability distribution of a
particle `behind the moon' may be instantaneously changed by an
arbitrarily small instantaneous change of the potential `inside the
laboratory'.
\end{quote}
Since the change of the potential may be arbitrarily small, such
superluminal effects are unacceptable in spite of the nonrelativistic
character of the theory.
\vskip 5mm

One might try to exclude relevant Gisin effects by changing the coupling
of nonlinear quantum mechanical systems. However, the coupling should be
\begin{itemize}
\item
the same for uncorrelated subsystems,
\item
invariant under nonlinear gauge transformations, and
\item
mathematically respectable for sufficiently many wave functions.
\end{itemize}
Unfortunately, no modification fulfilling these requirements is known up
to now.
\vskip 5mm

Of course, the results presented in this paper do not indicate any
problem for the general Doebner-Goldin equation if it is interpreted as a
1-particle equation, as originally suggested \cite{DoGo,DoGoGen}.
\medskip

\section*{Acknowledgement}
I am indebted to H.D.~Doebner for strong encouragement.

\end{document}